\documentstyle[prd,aps,twocolumn]{revtex}

\def\a{\alpha}

\def\e{\epsilon}

\def\th{\theta}

\def\del{\partial}

\def\bar#1{\overline{ #1 }}

\def\be{\begin{equation}}
\def\ee{\end{equation}}
\begin{document}

\title{$\theta$ Vacua in the Light-Cone Schwinger Model }

\author{Alex C. Kalloniatis\thanks{Present Address: Institute for
Theoretical Physics, University of Erlangen-N\"urnberg, 91058
Erlangen, Germany}} 
\address{Max-Planck-Institut f\"ur Kernphysik, D-69029 Heidelberg,
Germany}

\author{ David G. Robertson }
\address{ Department of Physics, The Ohio State University, 
Columbus, OH 43210, USA }

\date{February 1, 1996}

\maketitle

\begin{abstract}

\noindent ABSTRACT:
We discuss the bosonized Schwinger model in light-cone quantization,
using discretization as an infrared regulator.  We consider both the
light-cone Coulomb gauge, in which all gauge freedom can be removed
and a physical Hilbert space employed, and the light-cone Weyl
(temporal) gauge, in which the Hilbert space is unphysical and a
Gau{\ss} law operator is used to select a physical subspace.  We
describe the different ways in which the $\theta$ vacuum is manifested
depending on this choice of gauge, and compute the $\theta$-dependence
of the chiral condensate in each case.

\vspace{.2in}
\noindent Preprint No.: OSU--NT--95--07

\end{abstract}
\pacs{ }

\paragraph*{Introduction.}

The method of Discretized Light-Cone Quantization (DLCQ) \cite{PaB85}
has recently become a viable nonperturbative tool for studying quantum
field theories, especially in two space-time dimensions, but possibly
also in four \cite{overview}.  It neatly unites the advantages of an
infrared regulated framework and the vacuum simplicity of Dirac's
``front form'' of relativistic dynamics \cite{Dir49}, and has been
applied to a variety of toy models with considerable success.

The simplicity of the vacuum is a major advantage of the light-cone
approach \cite{Wei66}.  It is also a puzzle, however, particularly in
light of the nontrivial physics associated with, e.g., the QCD vacuum.
It is therefore important to understand how physics that is normally
related to the vacuum appears in the light-cone framework.  In DLCQ,
any vacuum structure must necessarily be connected with the $k^+=0$
Fourier modes of the fields.\footnote{This follows from simple
kinematical considerations.}  There has recently been a great deal of
effort devoted to studying the properties of these zero modes, which
can be quite nontrivial \cite{czm}.  This work has shown that some
types of vacuum structure---spontaneous breaking of discrete
symmetries in scalar field theories, for example---is in fact
recovered with a careful treatment of the zero modes.

The purpose of this note is to discuss the connection between zero
modes and another type of vacuum structure: the $\theta$ vacuum.  We
shall address this in the simplest nontrivial setting, namely the
Schwinger model \cite{schwing}.  This model has been discussed
extensively in the light-cone literature, mainly in the fermionic
representation.  McCartor in particular has given a thorough treatment
of the fermionic version \cite{McCartor}.  There are many subtleties
that must be addressed in order to understand the vacuum structure
from this point of view---the left-moving fermions, the proper
definition of operator products, and the selection of a suitable
physical subspace, to name a few.  The anomaly relation is also bound
up in the subtleties, and is closely connected to the vacuum
structure.

Here we shall sidestep most of these difficulties by studying the
bosonized form of the theory.  In this case the anomaly relation, for
example, is obtained directly as an equation of motion.  In addition,
the condition that physical states be chargeless is automatically
implemented through the bosonization.  The only nontrivial issue that
remains is the $\theta$ structure, which can thus be studied in
isolation.  We shall focus in particular on the $\theta$-dependence of
the chiral condensate $\langle \th|\bar{\psi}\psi |\th \rangle$ in
this model.  This represents the only ``observable'' consequence of
the $\theta$ vacuum \cite{schwing}.  We shall study this dependence in
two different gauges---the light-cone Weyl gauge, in which we have an
extended Hilbert space, and the light-cone Coulomb gauge, in which we
eliminate all gauge freedom at the classical level.  That the precise
manifestation of the topological structure can be gauge-dependent is
well known \cite{Raj}.  Our aim is to exhibit the gauge-dependence of
the $\th$-structure in light-cone quantization.  As we shall see, the
correct results are obtained, but in a somewhat different way than in
the conventional approach.

\paragraph*{Canonical Formalism.}

Our starting point is the bosonized form of the Schwinger model
Lagrangian
\begin{equation}
{\cal L} = \frac{1}{2} (\partial_\mu \phi)( \partial^\mu \phi)
- g A_\mu \epsilon^{\mu \nu} \partial_\nu \phi
- \frac{1}{4} F_{\mu \nu}F^{\mu \nu}\; ,
\label{lagrangian}
\end{equation}
where $F_{\mu \nu} = \partial_\mu A_\nu - \partial_\nu A_\mu$ and
$\epsilon^{\mu\nu}$ is completely antisymmetric with $\epsilon^{01} =
+1$.  Some useful correspondences with the fermionic version of the
model are $g = e/\sqrt{\pi}$, relating the mass of the Schwinger boson
to the gauge coupling $e$, and
\begin{equation}
\bar{\psi}\psi = K :\! \cos (2 \sqrt{\pi} \phi)  \!:\; ,
\label{chiralcorr} 
\end{equation}
where $K= e^{\gamma_E}$ with $\gamma_E$ the Euler-Mascheroni constant.

We choose the light-cone convention $x^\pm = (x^0\pm x^1)/\sqrt{2}$,
and quantize independent fields on the line $x^+=0$.  We take space to
be a finite interval, $-L\leq x^-\leq L$, with periodic boundary
conditions on the fields.  It is therefore important to distinguish
the zero and normal mode parts of fields in a Fourier expansion.  We
adopt the notation $\phi(x^-) = \phi_0 + \varphi(x^-)$ and $A^\mu(x^-)
= A^\mu_0 + A^\mu_n(x^-)$, with 
\be
\phi_0 = {1\over2L}\int_{-L}^{L} dx^- \phi 
\ee
and analogously for $A^\mu_0$. When a given expression holds for
both zero and normal modes, however, the distinction will be
suppressed. In addition, the periodic delta function with its zero
mode subtracted will be needed.  We define
\begin{equation}
\tilde{\delta} (x) \equiv \delta(x) - \frac{1}{2L} \; ,
\end{equation} 
where
\begin{equation}
\delta(x)={1\over2L}\sum_{n=-\infty}^\infty e^{in\pi x/L}
\end{equation}
is the full periodic delta function.

The canonical momenta are computed according to, for example,
\be
\pi_\phi = {\del{\cal L}\over \del(\del_+\phi)}\; .
\ee
We find
\begin{eqnarray}
\pi_{\varphi} & = & \partial_- \varphi - g A^+_n \\  
\pi_{\phi_0}  & = & - g A^+_0 \\
\pi_{A^+_n}   & = & \partial_+ A^+_n - \partial_- A^-_n \equiv \pi^-_n \\
\pi_{A^+_0}   & = & \partial_+ A^+_0 \equiv \pi^-_0 \\
\pi_{A^-}     & = & 0 \;.
\end{eqnarray}
Together with the canonical Hamiltonian,
\begin{equation}
P^- = \int_{-L}^L dx^- \left[ \frac{1}{2} (\pi^-)^2 - 
A^- \left(\partial_- \pi^-_n + g \partial_- \varphi\right)\right]\; ,
\end{equation}
this represents a system with both first and second class constraints
in the sense of Dirac \cite{Dir64}.  To determine the appropriate
quantum commutators we must first introduce gauge conditions and then
either pursue the Dirac-Bergmann program \cite{Dir64}, or seek to
implement the equations of motion correctly as Heisenberg equations.
We shall consider two different gauges: the light-cone Weyl (or
temporal) gauge, $A^- = 0$, and the light-cone Coulomb gauge,
$\partial_- A^+ = 0$.  It is easy to check that both gauges are
consistent with the periodic boundary conditions we have imposed in
the present light-cone formulation.

Details of the Dirac-Bergmann procedure have been given in
many places, including in the light-cone literature
\cite{MaY76,Hei92}, and we shall not repeat them here.  We shall
instead simply give the final field algebra and dynamical operators in
each of the two gauges.  In each case we then discuss the origin of
the $\theta$ structure, and compute the chiral condensate.

\paragraph*{ Light-Cone Weyl Gauge.}
 
Imposing the condition $A^-=0$, we find that the appropriate
commutation relations are
\begin{eqnarray}
\left[ \pi^-_n (x^-), \partial_- \pi^-_n(y^-) \right] & = &
\frac{i g^2}{2} \tilde{\delta} (x^- - y^-) \label{pipi} \\   
\left[ A^+_n(x^-) , \pi^-_n(y^-) \right] & = &  
i \tilde{\delta} (x^- - y^-)   \label{A+pi-} \\
\left[\partial_- \pi^-_n (x^-) , \varphi(y^-) \right] & = & 
 \frac{i g}{2} \tilde{\delta} (x^- - y^-)  \label{phipi} \\
\left[ \varphi(x^-), \partial_- \varphi(y^-)\right] & = &  
\frac{i}{2} \tilde{\delta} (x^- - y^-)  \label{phiphi} \\
\left[ g A^+_0 , \phi_0 \right] & = & \frac{i}{2L} \label{A+phi0}  
\;.
\end{eqnarray}
To these should be added the condition $\pi^-_0 = 0$, which arises as
a secondary constraint, and the Gau{\ss} law condition defining physical
states,
\begin{equation}
G |{\rm{phys}}\rangle = 0\; , \quad
G \equiv\left( \partial_- \pi^-_n + g \partial_- \varphi\right)
\; ,
\label{Gauss}
\end{equation}
which arises from the first-class constraint that remains after
imposing the condition $A^- = 0$.  The Hamiltonian is simply
\begin{equation}
P^- = \frac{1}{2} \int_{-L}^L dx^- \left(\pi^-_n\right)^2\; ,
\label{weylham} 
\end{equation}
which does not immediately reflect that the physical spectrum of the
theory is that of a free boson of mass $g$.  This is evident only
after one has satisfactorily implemented Eq. (\ref{Gauss}) and
identified the physical subspace.  That the correct spectrum is
obtained for physical states can be seen, for example, by rewriting
Eq. (\ref{weylham}) as
\begin{equation}
P^- = \frac{1}{2} \int_{-L}^L dx^- \left( g \varphi - 
\frac{1}{\partial_-} G \right)^2\; .
\end{equation}
Thus the correct physical spectrum is obtained in matrix elements
between physical states, $\langle {\rm{phys}}| P^-
|{\rm{phys}}^{\prime}\rangle$.

As usual, the Gau{\ss} operator, $G$, is the generator of residual,
i.e., $x^+$-independent, gauge transformations.  A finite gauge
transformation of this type is implemented by the unitary operator
\be
\hat{U}[\omega] = \exp \left(i\int_{-L}^L dx^-\omega(x^-)G(x^-)
\right) \; ,
\label{smallgt}
\ee
with
\be
\hat{U}[\omega] A^+ \hat{U}^\dagger[\omega] = A^+ + \del_-\omega\; .
\ee
For periodic gauge functions $\omega$, physical states satisfy
\be
\hat{U}[\omega]|{\rm phys}\rangle  = |{\rm phys}\rangle\; ,
\ee
that is, they are invariant with respect to these residual gauge
transformations.  There exist gauge transformations, however, that are
not themselves periodic yet still preserve the periodic boundary
condition on the gauge field.  These ``large'' gauge transformations
may be decomposed into a product of a small transformation
(\ref{smallgt}) and a transformation of the form
\begin{equation}
U_n=e^{in\pi x^-/L}\; ,
\label{largegt}
\end{equation}
where $n$ is any integer.  This specific structure is a consequence of
the form of the symmetries of the original fermionic theory.  The
transformation (\ref{largegt}) merely shifts the zero mode $A^+_0$:
\be
\delta A^+_0 = {1\over ie}U_n\del_-U_n^* = -{n\pi\over eL}\; .
\label{winding}
\ee
It will prove convenient to introduce the dimensionless field
\begin{equation}
z = \frac{ e A^+_0 L }{\pi}\; ,
\end{equation}
in terms of which Eq. (\ref{largegt}) takes
\begin{equation}
z\rightarrow z - n\; .
\end{equation}
The Gau{\ss} law condition, Eq. (\ref{Gauss}), does not require the
equivalence of physical states related by such a gauge transformation.

The situation is precisely analogous to that of equal-time
quantization in the temporal gauge $A^0=0$.  Physical states are
invariant under the residual transformations obtained by
exponentiating the Gau\ss\ operator, but only phase-invariant under
the large gauge transformations analogous to (\ref{largegt}).  That
is, if $\hat{U}_n$ is the unitary operator that implements the
transformation defined by (\ref{largegt}), then
\begin{equation}
\hat{U}_n|{\rm phys}\rangle = e^{-in\theta}|{\rm phys}\rangle\; .
\label{uaction}
\end{equation}
The specific form of the phase factor follows from the need to respect
the composition law $\hat{U}_n \hat{U}_m = \hat{U}_{n+m}$.

In order to discuss the chiral condensate let us give a specific
realization of the $\theta$ states.  Since $z$ commutes with the other
fields, any state in the theory can be written as a superposition of
states of the form
\be
\psi_\theta(z)\otimes \Phi[A^+_n,\varphi]\; .
\ee
The state $\Phi[A^+_n,\varphi]$ can be thought of as either in a Fock
or a functional Schr\"odinger representation, and must be annihilated
by the Gau\ss\ operator in order to be in the physical subspace.  The
zero mode wavefunction $\psi_\theta(z)$ is chosen to be an eigenstate
of $\hat{U}_n$ with eigenvalue $e^{-in\theta}$.  An explicit
representation for $\hat{U}_n$ is given by
\be
\hat{U}_n = e^{-inp_z}\; ,
\label{uhat}
\ee
where $p_z$ is the momentum conjugate to the rescaled variable $z$:
\be
p_z\equiv 2\sqrt{\pi}\phi_0\; ,
\ee
so that 
\be
[z,p_z] = i\; .
\label{zpzcomm}
\ee
That Eq. (\ref{uhat}) is correct may be seen from
\be
\hat{U}_n z \hat{U}_n^\dagger = z-n\; .
\ee
In this coordinate representation, $p_z$ is represented as a
derivative operator
\be
p_z = -i{d\over dz}\; ,
\ee
and a convenient choice for the state $\psi_\theta(z)$ is
\be
\psi_\theta(z) = e^{i\theta z}\; .
\ee
Note that this state is not, strictly speaking, normalizable.  It
does, however, satisfy the usual orthogonality relation
\begin{eqnarray}
\langle \theta|\theta^\prime \rangle & = &
\int_{-\infty}^\infty dz\, \psi_\theta^*(z)\psi_{\theta^\prime}(z)
\nonumber\\
& = & \delta (\theta - \theta^\prime )\; .
\end{eqnarray}

With these explicit expressions for the states and operators in hand,
we can now compute the chiral condensate
$\langle\theta|\bar{\psi}\psi|\theta\rangle$.  The physical $\th$
vacuum state will be a tensor product of $\psi_\theta(z)$ with the
Fock vacuum for $A^+_n$ and $\varphi$.  Using the correspondence
formula Eq. (\ref{chiralcorr}), and the fact that the normal mode part
of the vacuum is the Fock vacuum, we find that only the scalar zero
mode contributes:
\begin{eqnarray}
\langle \theta | \bar{\psi}\psi | \theta \rangle
&=& K \int_{-\infty}^\infty dz\,
\psi_\theta^*(z) :\! \cos (2\sqrt{\pi}\phi_0) \!: 
\psi_{\theta^\prime}(z) \\
&=& K \int_{-\infty}^\infty dz\,
\psi_\theta^*(z) \left({e^{ip_z}+e^{-ip_z}\over2}\right)
\psi_{\theta^\prime}(z)\; .
\end{eqnarray}
Making use of Eq. (\ref{uaction}), and dividing out the normalization
factor, we obtain the standard result
\begin{equation}
\langle \theta | \bar{\psi} \psi | \theta \rangle= K \cos\theta\; .
\end{equation}

Note that the only place $p_z$ appears in the theory is in the
operator that implements large gauge transformations.  In particular,
there is no contribution from the zero mode sector to the Hamiltonian.
This is actually unique to the light cone. In the equal-time
formulation the two sectors decouple in the Hamiltonian
\cite{schwing}.  Nevertheless, the standard result follows: the
occurrence of the $\th$ vacuum has no effect on the spectrum or other
physical properties of the theory.  Note also that the physically
distinct values of $\th$ lie in the range $0\leq \th \leq 2\pi$, again
in accordance with the standard results \cite{schwing}.

\paragraph*{ Light-Cone Coulomb Gauge.}

The gauge $\partial_- A^+=0$ is in many ways the most natural
one for this problem in that there is no residual first-class
(Gau{\ss} law) constraint to be imposed on the states.  Rather, the
latter can be used to eliminate $A^-$ at the operator level.  (The
zero mode of $A^-$ may be set to zero by a purely $x^+$-dependent
gauge transformation.)  Moreover, the Hamiltonian is just that of a
free massive boson,
\begin{equation}
P^- = {{g^2}\over{2}} \int_{-L}^L dx^- \varphi^2 \; ,
\end{equation} 
so that the physical spectrum of the theory is manifest.  The relevant
commutators are
\begin{eqnarray}
\left[ \varphi(x^-), \partial_- \varphi(y^-)\right] & = &  
\frac{i}{2} \tilde{\delta} (x^- - y^-) \\
\left[ g A^+_0 , \phi_0 \right] & = & \frac{i}{2L}
\; ;
\end{eqnarray}
all other variables are eliminated by constraints.  These commutators
were first derived in Ref. \cite{Hei92}.  Once again, the scalar field
zero mode is the momentum conjugate to the gauge zero mode.

The light-cone Coulomb gauge condition, however, does not completely
fix the gauge, due to Eq. (\ref{winding}).  The large gauge
transformations connect different Gribov regions \cite{gri78}.  We can
eliminate this remaining gauge freedom by restricting $z$ to lie in a
``fundamental modular domain,'' for example $0\leq z\leq1$ (with the
points $z=0$ and $z=1$ identified).  This uses up all remaining
freedom and completely fixes the gauge.

Again, a general state can be represented as a superposition of states
of the form $\psi(z) |{\rm Fock}\rangle$, where the Fock state is
constructed from the modes of $\varphi$.  Here, without loss of
generality, the function $\psi(z)$ may be taken to be periodic on the
fundamental domain.  A convenient representation in terms of a
complete orthonormal set is
\begin{equation}
\Psi_n(z) ={1\over\sqrt{2L}} e^{2in \pi z} \;.
\label{wavefns} 
\end{equation} 
The momentum operator $p_z$ is again represented as a derivative, but
the representation is not unique; the most general realization of the
commutator (\ref{zpzcomm}) takes the form
\begin{equation}
p_z = -i \frac{d}{dz} + \theta\; .
\end{equation}
Note that $\th$ can be shifted out of $p_z$ and into the states by the
transformation
\be
\xi_n^\th(z) = e^{i\th z}\Psi_n(z)\; .
\label{newstates}
\ee
This new state satisfies the boundary condition $\xi_n^\th(0) = e^{i\th}
\xi_n^\th(1)$, and the transformed momentum operator is simply
\be
\pi_z = -i{d\over dz}\; .
\end{equation}
Because the zero modes decouple completely from the normal modes in
$P^-$, the physical vacuum will be the product state
\begin{equation}
|\th \rangle = \xi_n^\th(z) |0\rangle\; ,
\end{equation}
with $|0\rangle$ the light-cone Fock vacuum of the scalar field and
$\xi_n^\th$ any one of the wavefunctions of Eq. (\ref{newstates}).

The condensate is now easily evaluated in the same way as before.
Again, only the scalar field zero mode contributes:
\begin{equation}
\langle \th | \bar{\psi} \psi | \th \rangle
= K \int_0^1dz\, \xi_n^\th(z)^* :\! \cos (p_z) \!: \xi_n^\th(z)\; .
\end{equation}
Expressing $\phi_0$ in terms of $\pi_z$ and expanding the cosine then
gives
\begin{eqnarray}
\langle \th | \bar{\psi} \psi | \th \rangle
&=& K \cos (2n\pi + \theta)\nonumber\\
&=& K \cos\theta\;.
\end{eqnarray}

As before, the zero mode operators do not appear in the Hamiltonian so
that the value of $\th$ has no effect on the spectrum of the theory.
In addition, only the values $0\leq\th\leq2\pi$ are physically
distinct, as expected.

\paragraph*{ Chiral Transformations.}

In the bosonized theory the chiral current is given by
\be
J^\mu_5 = {1\over\sqrt\pi}\del^\mu \phi\; .
\ee
The correct anomaly relation for $J^\mu_5$ follows directly from the
equation of motion for $\phi$:
\begin{eqnarray}
\del_\mu J^\mu_5 &=& {1\over\sqrt\pi}\del_\mu\del^\mu\phi
\nonumber\\
&=& -{e\over\pi}\e_{\mu\nu}F^{\mu\nu}\; ,
\end{eqnarray}
so that this aspect of the model is automatic in the bosonized
version.  In the fermionic theory the anomaly is quite nontrivial,
involving a range of subtleties regarding the definition of singular
operator products.

A conserved axial current can be defined, however, and is given by
\be
J^\mu_{5,sym} \equiv {1\over\sqrt\pi}\left(\del^\mu \phi
+g\e^{\mu\nu} A_\nu\right)\; .
\ee
$\del_\mu J^\mu_{5,sym} = 0$ reflects the invariance of the theory
under shifts in $\phi$.  The associated symmetry charge is
\begin{eqnarray}
Q_{5,sym} &=& \int^L_{-L} dx^- J^+_{5,sym}\nonumber\\
&=& -2z\; ,
\end{eqnarray}
In both of the gauges we have discussed, this charge generates
transformations that change the value of $\th$:
\be
e^{i\alpha Q_{5,sym}}|\th\rangle = |\th-2\a\rangle\; .
\ee
This is the only effect of a chiral transformation in the theory
\cite{schwing}.

\paragraph*{ Discussion.}

We have seen that in light-cone quantization the $\th$ vacuum
structure of the bosonized Schwinger model can be reproduced by a
careful treatment of the zero momentum modes of the fields defined on
a compact space.  The precise manner of its manifestation is somewhat
gauge-dependent, as is familiar from the equal-time formulation.  In
the LC temporal gauge, one works in an extended Hilbert space and the
residual gauge freedom is removed by identifying only those states
that are annihilated by the Gau\ss\ operator as physical.  The $\th$
structure enters because the Gau\ss\ condition does not enforce
gauge-equivalence of states related by certain ``large'' gauge
transformations.  These states are only phase invariant, with $\th$
being the arbitrary phase that enters the transformation rule.  This
is precisely analogous to the equal-time formulation in the gauge
$A^0=0$.

In contrast, the LC Coulomb gauge formulation is physical a one, in
the sense that all gauge freedom can be removed at the classical level
and a purely physical Hilbert space employed.  It is natural to do
this and work in a finite ``fundamental modular domain'' for the gauge
field zero mode.  In this case, $\th$ enters as an arbitrariness in
the representation of the conjugate momentum $p_z$ as a derivative,
or, equivalently, as an arbitrariness in the boundary condition
satisfied by the zero mode wavefunction on the fundamental domain.
Again, this is quite familiar in the analogous equal-time context.

In either gauge the expected features of the model are reproduced,
although these are rather simple.  The spectrum of the theory in each
case is that of a free boson of mass $g=e/\sqrt{\pi}$, and is
independent of the value of $\th$.  The only quantity that is
sensitive to the value of $\th$ is the chiral condensate, and its
$\th$-dependence is correctly obtained.  The crucial feature in each
case is the presence of a vacuum wave function with the structure
$\psi(z)\sim e^{i\th z}$, along with the fact that the zero mode of
the scalar field, which appears in the bosonized expression for
$\bar{\psi}\psi$, is the momentum conjugate to the variable $z$.  In
the LC temporal gauge, the necessary vacuum wave function arises
because states need only be phase-invariant under ``large'' (residual)
gauge transformations.  The role of the zero mode wave function is to
supply this phase when acted on by the appropriate unitary operator.
In the LC Coulomb gauge, a $\th$-dependent boundary condition on the
fundamental domain is permissible, which leads to a similar structure
in the zero mode wave function.

In the presence of a fermion mass, the $\th$ vacuum has a definite
impact on the spectrum of QED$_{1+1}$.  That we obtain the correct
results for the massless case, albeit in the bosonized form of the
model, gives a reasonable basis for the extension to massive fermions.

\acknowledgments

DGR thanks the Max-Planck-Institut f\"ur Kernphysik, Heidelberg, for
its kind hospitality during the early stages of this work.  ACK was
supported by a Fellowship from the {\it Max-Planck Gesellschaft}.  DGR
was supported by the National Science Foundation under Grants
Nos. PHY-9203145, PHY-9258270, and PHY-9207889, and in the final phase
of this work by a grant from the US Department of Energy.  Travel
support was provided in part by a NATO Collaborative Grant.

\end{document}